\documentclass[final,5p,times,twocolumn]{elsarticle}

\usepackage{amssymb}
\usepackage{graphicx}  % needed for figures
\usepackage{dcolumn}   % needed for some tables
\usepackage{bm}        % for math
\usepackage{amssymb}   % for math
\usepackage{amsmath}
\usepackage{empheq}
\usepackage{graphicx,color}

\def\ket#1{| #1 \rangle}
\newcommand{\unit}{1\!\!1}

\journal{Physics Letters A}

\begin{document}

\begin{frontmatter}



\title{Coarse Quantum Measurement : an analysis of the Stern-Gerlach Experiment}


\author{Anirudh Reddy
$^{1}$,  Joseph Samuel$^{1}$, Kumar Shivam$^{1}$\ and
Supurna Sinha$^{1}$}

\address{$^{1}$Raman Research Institute, Bangalore 560080, India.}

\begin{abstract}
We study the Quantum Measurement Process in a Stern-Gerlach setup 
with the spin of a silver atom as the quantum system and the position 
as the apparatus. The system and the apparatus are treated quantum-mechanically 
using unitary evolution. The new ingredient in our analysis is the idea that the 
probes determining the position of the silver atom are limited in resolution. 
We show using a Wigner matrix that due to the coarseness of the detection process, 
the pure density matrix appears to evolve to an impure one. We quantify the information 
gained about the spin in a coarse position measurement.
%%\rev{(where, $f=\frac{maT}{\hbar}(\xi+\frac{1}{3}aT^2)$)}

\end{abstract}


\end{frontmatter}


\section{Introduction}
\label{}
Quantum mechanics is a very successful theory for describing the microscopic world of atoms.
However, ever since its inception there are certain fundamental aspects of quantum theory 
that have remained
obscure. This has to do with the relation between unitary evolution which is central to the 
theory, and the measurement
process, which gives us information about the quantum system. 
The challenge that remains is a self consistent formulation of quantum theory which explains unitary evolution and outcomes
of a measurement within a single framework.

Bohr had taken a semiclassical approach in which
he viewed the apparatus classically and treated the system (spin)
quantum mechanically. Such a point of view is unsatisfactory since at a fundamental level the world is 
governed by quantum mechanics.
In this letter, we present a completely quantum mechanical analysis of the Stern-Gerlach experiment. 
Our purpose is to explore, in a simple solvable context, the idea
that coarseness of the experimental probes is responsible for {\it apparent}
non unitarity in the measurement process.

We focus on the Stern-Gerlach experiment as a
context for understanding the measurement process in quantum mechanics
without invoking any {\it ad hoc} assumption beyond pure unitary evolution.
Let us begin by summarizing the Stern-Gerlach experiment.
The set up consists of a beam of silver atoms
(spin-${\frac{1}{2}}$ particles) moving along the $z$ direction passing through an
inhomogeneous magnetic field along the $y$ direction. Two spots appear on
the screen corresponding to the $y$ component of the spin, $S_y=\frac{1}{2}$ and $S_y=-\frac{1}{2}$.
There have been a few analytical studies of this experiment 
in the past couple of decades\cite{anua,anub,gondran,widom,platt,berman,frasca,scully}.
%In some of these studies\cite{gondran,platt,widom} the authors arrive at the final observation of the appearance of two discrete spots in a 
%Stern-Gerlach 
%experiment by considering a time averaging. 
%The focus of some of them is to establish quantization of angular momentum \cite{lieberman}.
Some studies\cite{frasca}
invoke Ehrenfest's Theorem to address the issue of measurement in a Stern-Gerlach setup.    
There have also been detailed analyses of the Stern-Gerlach experiment from the point of view of 
environment induced decoherence\cite{anua,anub,berman,gomis}. 
%There are some other analyses\cite{scully} which use a magnetic field 
%satisfying  
%Maxwell's equations, correcting for earlier analyses which used magnetic fields violating Maxwell's equations. 
In this letter we invoke the new idea that there is an inherent {\it coarseness} in the detection process. 
The role of coarseness of the measurement process in the quantum to classical transition 
has been explored in the past\cite{kb,jlk,rss}. In \cite{kb,jlk} coarseness of the measurement process has been investigated by using the 
Leggett-Garg inequality as a way of probing the quantum to classical transition. In \cite{rss} coarse graining has been used as a probe for 
detection of entanglement between a microscopic system and a macroscopic system. We present an analysis which offers a new perspective on the 
Stern-Gerlach experiment from the point of view of the coarseness of the measurement process. 

The letter is organized as follows. We first summarize the measurement process. Then we present
a theoretical analysis of the Stern-Gerlach experiment 
using unitary evolution. We show with a Wigner function matrix approach, that a coarse measurement 
with a finite spatial resolution leads to an apparently non unitary evolution for the Wigner matrix.
Finally we end our letter with a concluding discussion. 
%*******************
%Figures to be displayed:\\
%2. Figure 2 displays a family of snapshots of the interference pattern at various values of $z$. Fig 3 is a zoomed in 
%version of one of the figs in Fig 2. Notice the reappearance of fringes on zooming in.
%1. Figure 1 displays a schematic SG expt.
%*****************************

\section{The Measurement Process}
\label{}
Let us summarize the measurement process in quantum mechanics. 
Our system is initially in a coherent superposition of states
$|S\rangle=\sum_i c_i |S_i\rangle$ in an orthonormal basis which 
diagonalises the quantity being measured.
To begin with, the system plus apparatus
is in  the product state $|\psi\rangle=|S\rangle|A\rangle$, 
in which the system and apparatus are unentangled. 
It is useful to logically break up the measurement process into three steps.
The first step in the measurement process entails 
coupling between the quantum system and the measuring apparatus so that the total state evolves unitarily to an entangled state 
$U|{\psi}\rangle = \sum_{i}c_i|S_i\rangle|A_i\rangle$\footnote{We note that in general, the $|A_i\rangle$ s need not be orthonormal.}. 
This state can be represented as a pure density matrix 
\begin{equation}
\rho=|\psi\rangle\langle\psi| = \sum_{ij}c^*_jc_i|S_i\rangle|A_i\rangle \langle S_j| 
\langle A_j|
\end{equation}
After the second step, the density matrix of the system takes the impure form
\begin{equation}
\tilde{\rho}= \sum_{i}|c_i|^2|S_i\rangle\langle S_i|
\label{impure}
\end{equation} 
which is interpretable as a classical mixture of states. 
Finally, the impure diagonal density matrix (\ref{impure}) 
goes over to a pure state
$c_i|S_i\rangle\langle S_i|c_i^{*}$.
The first step can be explained entirely in terms of unitary evolution 
and therefore is not controversial. 
The final step, sometimes called ``collapse", has been debated extensively as the ``quantum measurement problem". This singling out of 
one outcome from many possibilities is not addressed here. Let us note that, even in classical probability theory, there is a
singling out of one from several outcomes (only one horse wins the race).  
We address here the second step; the transition from quantum superpositions to 
classical mixtures. This is the focus of our letter.  
In this letter, we investigate the Stern-Gerlach experiment from 
the perspective of
coarse quantum measurement (CQM), 
in which 
we recognize the fact that all experiments
are constrained by bounded resources. We model these constraints by using a screen 
whose size and spatial resolution are fixed. The spatial resolution of the screen is given by
the pixel size and the size of the screen determines the total number of pixels. Experimentally
one can only say that an atom was incident on our screen somewhere within a pixel. 
Fixing these resources imposes ultraviolet as well as infrared cutoffs on the experimental probes. 
In this letter we are more concerned with the short distance cutoff.

\section{Stern-Gerlach}
\label{}
Consider silver atoms with spin $\frac{1}{2}$ 
at rest in the laboratory, 
%%along the $z$ direction,  
in a magnetic field given by 
$$B=(B_0 y,B_0 x,0).$$ Notice that this field is both divergence and curl free. We confine the atoms to the $x$-$z$ 
plane and thus set $y=0$. 
The Hamiltonian for the system 
is 
\begin{equation}
H=\frac{p^2}{2m}+Fx\sigma_y
\label{hamil}
\end{equation}
where $$F=-g\mu_B\frac{\hbar}{2}B_0,$$
with $g$ the Land\'{e} $g$ factor and $\mu_B$ the Bohr magneton.  
The Schr{\"o}dinger equation for the system is given by(See Appendix A)
\begin{equation}
i\hbar\frac{\partial{\psi}}{\partial t}=-\frac{ \hbar^2}{2m}{\nabla^2 \psi}+Fx\sigma_y\psi,
\label{evol}
\end{equation}
%$|\psi>=
%\left(
%\begin{array}{c}
%\psi_+\\
%\psi_-\\
%\end{array}
%\right)$
where $\psi$ is a two component Pauli spinor. 
This problem can be solved exactly by transforming from the laboratory frame to a freely falling 
frame $\xi=x-\frac{1}{2} a t^2$ and $T=t$, with $a=F/m$ the acceleration and 
$\psi = e^{if} \phi$ (where, $f=\frac{maT}{\hbar}(\xi+\frac{1}{3}aT^2)$), (which reduces it to a free particle problem as in the Einstein elevator) and then transforming back to the laboratory frame\cite{steiner}(See Appendix A). 
Such an analysis is entirely quantum mechanical. We view the spin as a quantum system and 
the position of the silver atom as the apparatus or pointer. We do not invoke any semiclassical approximation. 
The formulation above results in a separation in {\it time} of the two spin states. 
This can be mapped to the formulation(See Appendix A), of a typical experiment, where 
the separation of the spins happens in {\it space} and we will sometimes use the spatial 
notation and language with the understanding that $t=zm/k\hbar$, 
where $\hbar k=\sqrt{2mE}$.
We start with an initial Gaussian wave packet of width $\sigma$. 
Setting $t=t_f-t_i$ 
our  analysis gives the following exact propagators
$K^{++}(x,x_i;t)$, $K^{--}(x,x_i;t)$ and $K^{+-}(x,x_i;t)$\cite{hsu,feynhibbs}(See Appendix A),
where the symbols $+$ and $-$ refer to the two components $\phi_+$ and $\phi_-$ of the 
Pauli spinor $\phi$.

\begin{eqnarray}\nonumber
K^{++}(x,x_i;t)&=&\label{}\sqrt{\frac{m}{2\pi i \hbar t}} {\exp}i[m\frac{(x-x_i)^2}{2 \hbar {t}}\\
& &-\frac{Ft(x+x_i)}{2 \hbar}-\frac{F^2t^3}{24m \hbar}]
\label{kernelplus}
\end{eqnarray}

\begin{eqnarray}\nonumber
K^{--}(x,x_i;t)&=&\label{}
\sqrt{\frac{m}{2\pi i \hbar t}} {\exp}i[m\frac{(x-x_i)^2}{2 \hbar t}\\
& &+\frac{Ft(x+x_i)}{2 \hbar }-\frac{F^2t^3}{24m \hbar}]
\label{kernelminus}
\end{eqnarray}

\begin{equation}
K^{+-}(x,x_i;t)=0
\label{kernelpm}
\end{equation}
The final wave function is got by ``folding'' the initial Gaussian with the propagator matrix (Eqs.\ref{kernelplus}-\ref{kernelpm}).
It has the form 
\begin{equation*}
\phi (x, t) = c_+\phi_{+} (x,t) |+\rangle + c_-\phi_{-} (x,t) |-\rangle,
\end{equation*}
where $|+\rangle$ and $|-\rangle$, 
are eigenstates 
of $\sigma_y$ and $\phi_+(x,t)$ and $\phi_-(x,t)$ are Gaussian wave packets (See Appendix A).
We can identify two relevant time scales (Eqs. A-16 and A-17 of Appendix A):
$\tau_1=\sqrt{\frac{2\sigma}{a}}$, the time over which the centers of mass of
the two wave packets separate and $\tau_2=\frac{m\sigma^2}{\hbar}$, the timescale
over which the individual wavepackets spread. We use the values
$m=1.79\times10^{-25} kg$, $F=9.27 \times 10^{-22} N$ and $\sigma=10^{-6}m$ which are 
experimentally reasonable.
Typical values for the two time scales are $\tau_1=10^{-5}s$ and
$\tau_2=10^{-3}s$.

We restrict our discussion to a situation where the detection screen is placed at a location just at the point where
the atom exits the magnetic field. However, in general one can have a further free evolution of the separated wavepackets
beyond this region in the field free space (See Appendix).

The full density matrix (See Appendix B) is of the form:
\begin{equation}
\rho_{\alpha\beta} (x, x^{\prime}) = 
c_{\alpha}\phi_{\alpha} (x) c_{\beta}^{*}{\phi}^{*}_{\beta} (x^{\prime}),
\label{density}
\end{equation}
where $\alpha$ and $\beta$ take values $+$ and $-$. In Eq.(\ref{density}) we have suppressed the time dependence in the notation.

%From the exact propagator, we isolate the two time scales.

%segna

\section{Entanglement and Coarse Graining}
\label{}
The total density matrix of the system has entanglement between the 
spin and atomic position.  The degree of entanglement can be measured
by the entanglement entropy \cite{Horo}, which is most easily computed by 
tracing over the position and diagonalising the $2\times2$ reduced density
matrix $\rho_{spin}$ for the spin. The result is
\begin{eqnarray}\nonumber
S_{ent}&=&-Tr [\rho_{spin} \log\rho_{spin}]\\
&=&\label{}\log{2}-\frac{(1+A(t))}{2}\log{(1+A(t))} \nonumber\\
& &-\frac{(1-A(t))}{2}\log{(1-A(t))}
\label{ententropy}
\end{eqnarray}
where $A(t)=\exp{-\frac{t^2(t^2+\tau_2^2)}{\tau_1^4}}$.
Here and below, we set $c_+=c_-=1/\sqrt{2}$ for simplicity. The entanglement
entropy is plotted in Fig. 1, which shows that $S_{ent}(t)$,
starts from zero at $t=0$, then increases and finally 
settles down to an asymptotic value  of $\log{2}$
over a time scale of the order of $10^{-7}s$.  This entanglement 
timescale is given by $\tau_3=\tau_1{}^2/\tau_2$ and is
shorter than the separation or spreading timescales. 
For $\tau_1\sim t>>\tau_3$, the entanglement
is high even though the wavepackets have not cleanly separated in real space(Fig. 2). 

From the density matrix $\rho_{\alpha\beta}(x,x')$ we construct the Wigner matrix $\mathbf{W}(q,p)$ \cite{gomis},
using the standard variables 
$q= (x+x^{\prime})/2$ and $y= (x-x^{\prime})$. The matrix elements of $\mathbf W(q,p)$ are: 
\begin{equation}
W_{\alpha \beta}(q,p)=\frac{1}{2\pi\hbar}\int_{-\infty}^{+\infty}{\rho_{\alpha \beta}(q+y/2,q-y/2) e^{\frac{ipy}{\hbar}} dy}
\label{wigmat}
\end{equation}
with $\alpha$, $\beta =\pm$.
$\mathbf{W}(q,p)$ is a $2\times2$ Hermitean matrix (not necessarily positive). 
All components of $\mathbf{W}(q,p)$ can {\it in principle} be measured by
having a Stern-Gerlach setup at the screen to measure
$Tr[\mathbf{W}(q,p)(\unit+\hat{n}.\vec{\sigma})/2]$.
In Figs $3$ and $4$, we display the function 
$W(q,p)={\rm Tr}[\mathbf{W}(q,p)(\unit+\sigma_x)/2]$, which shows 
the diagonal as well as off diagonal terms in the $\ket{+},\ket{-}$ basis.

\begin{figure}
\includegraphics[scale=0.6]{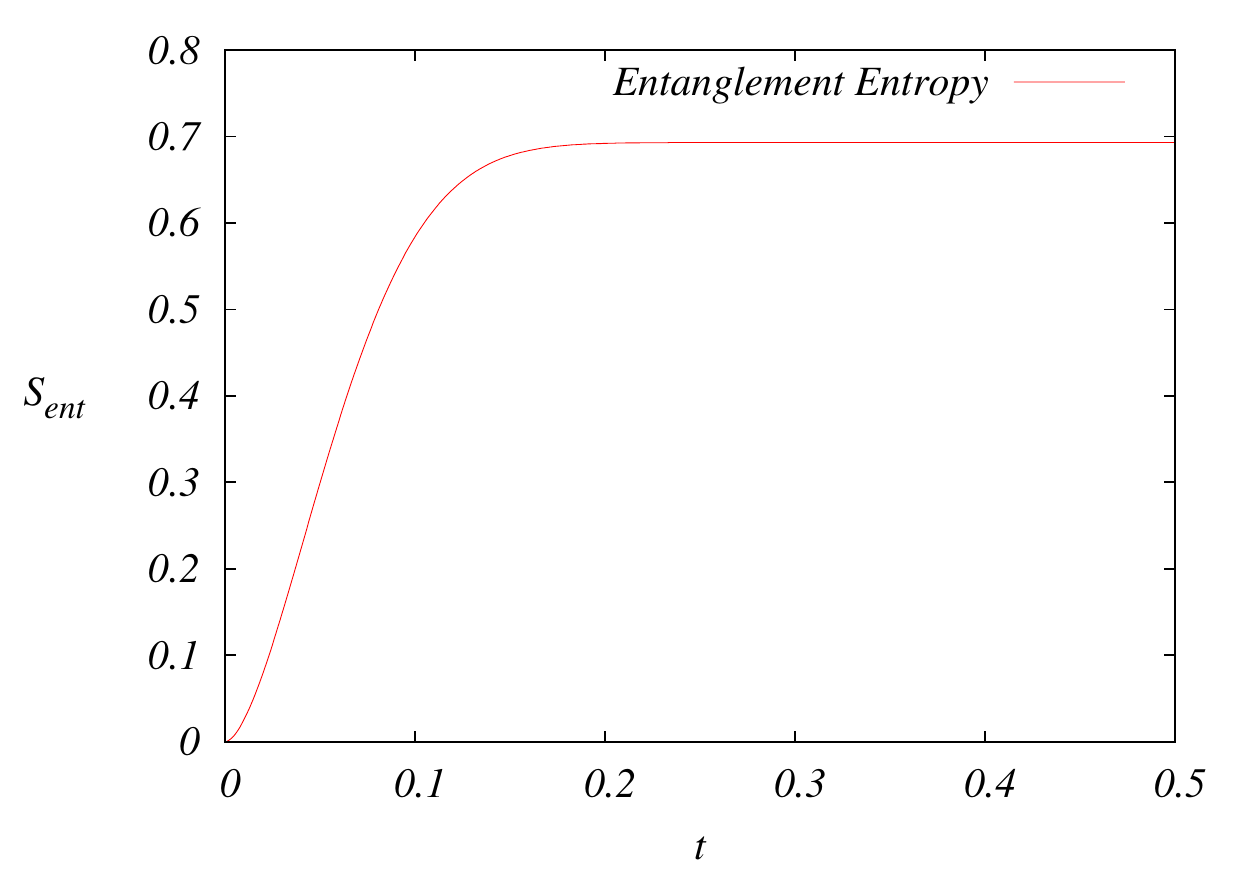}
\caption{\label{fig:epstart}
Figure shows the entanglement entropy as a function of time 
in units of $\mu s$. }
\end{figure}

\begin{figure}
\includegraphics[scale=0.6]{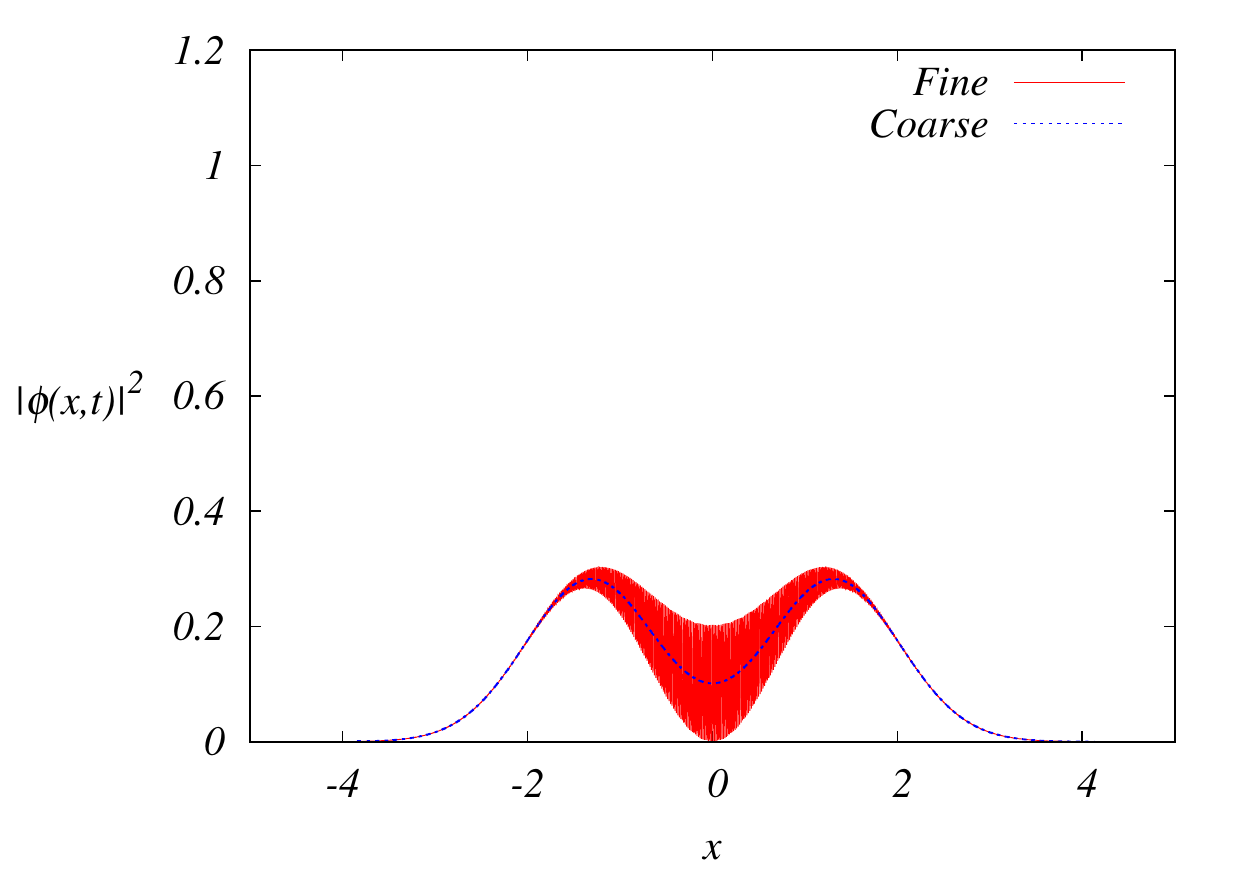}
\caption{\label{fig:epstart}
Figure shows the spatial distribution of the wavepacket at t=22.5$\mu$ sec.
We have used the typical values 
$m=1.79\times10^{-25} kg$, $F=9.27 \times 10^{-22} N$, $\sigma=10^{-6}m$.  }
\end{figure}

We now use the fact that the detection is done {\it coarsely}: the phase space resolution is poor and
so we integrate the Wigner matrix over volumes of 
phase space which are large compared to $h$.
The coarse grained Wigner matrix $\overline{\mathbf{W}} (q,p)$ (See Appendix B)
has elements: 
\begin{equation}
\overline{W}_{\alpha\beta} (q,p)=\frac{1}{\Delta \delta}\int_{-\Delta/2}^{\Delta/2} du\int_{-\delta/2}^{\delta/2} dv {W}_{\alpha\beta} (q+u,p+v) ,    
\label{wigneraverage}
\end{equation}
with $\Delta$ and $\delta$, the pixel size in position and momentum respectively. 
The off diagonal term ${W}_{+ -}(q,p)$ is oscillatory 
due to a term $e^{iq 2\pi/d}$, which oscillates on a length scale $d=\frac{\hbar}{2Ft}$(See Appendix B),
which is about $10^{-8} m$. On a coarse scale these off-diagonal elements average 
to zero and we have a diagonal matrix of the form(See Appendix B):
\begin{center}
$\left(\begin{array}{ccc}
\overline{W}_{++}(q,p) & 0\\
& & \\
0 & \overline{W}_{- -}(q,p)
\end{array}\right)$
\end{center}
%%\includegraphics[scale=0.3]{W0.pdf}
%%\caption{\label{fig:epstart}
%%Figure shows a plot of the Wigner function $W(x,p,t)$ as a function of $x$ and $p$ at $t=0$. 
%%We have chosen $\hbar =1$, $m=3$, $a=3$ and $\sigma=1$.} 
%%\end{figure}

What can one learn from a coarse measurement? 
Let us suppose as is usual in experiments that we only detect the position of the silver atom with low resolution 
and do not measure the momentum at all. We integrate the Wigner matrix over all momenta and integrate over a pixel
to find that
if an atom is detected at pixel location $X$, we would assign relative probabilities 
$P_{\pm}(X)=1/2\int_{-\Delta/2}^{\Delta/2}|\phi_{\pm}(X+u)|^2du $ and  
to its being spin $\pm$. By detecting an atom at pixel $X$ we do gain
information about the spin. If we set,
$P(X)=P_+(X)+P_-(X)$ and define $q_{\pm}(X)=P_{\pm}(X)/P(X)$, the entropy of the spin probability distribution is 
$S(X)=-q_-(X)\log{q_-(X)} -q_+(X)\log{q_+(X)}$. The information we gain is so given by
\begin{equation}
I(X)=\log{2}-S(X)
\label{info}
\end{equation} 
per event at $X$. Note that the arrivals at $X$ values away from $0$ give us more information.
The mean information per event is given by \cite{KLinfo}:
\begin{figure}
\includegraphics[scale=.3]{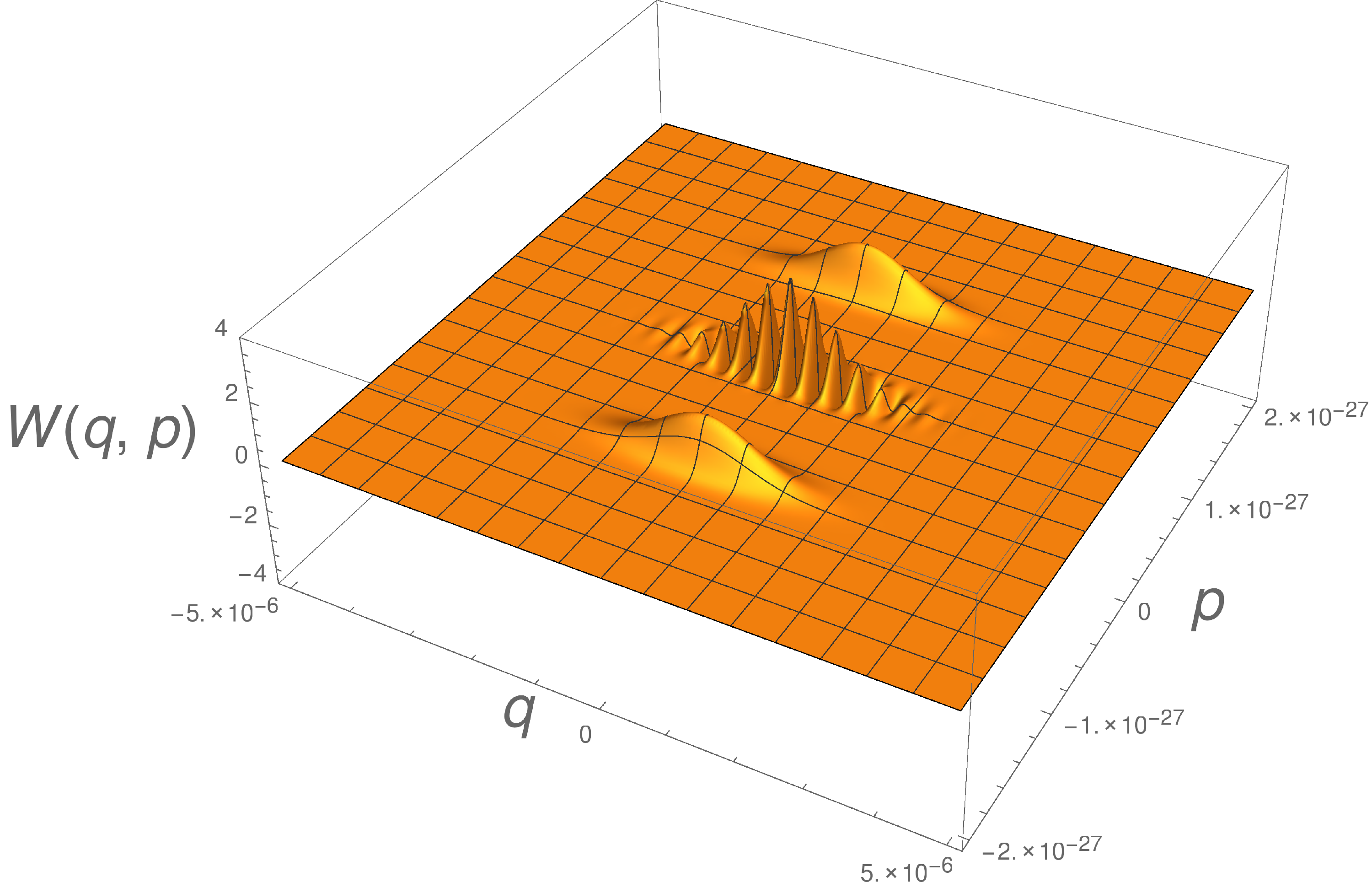}
\caption{\label{fig:epsart} 
Figure shows a plot of the function $W(q,p)$ (defined in the text) as a function of $q$ and $p$ 
at $t=1\mu$ sec. The central hump showing oscillations is the real part of the off-diagonal element of the Wigner matrix and
the others are diagonal elements. The remaining parameters and axes units are as mentioned in the caption of Fig. 2. The $W(q,p)$ axis has been rescaled by multiplying by a factor of $10^{-33}$. $q$ is displayed in $m$ and $p$ in $kg m/s$. }
\end{figure}

\begin{figure}
\includegraphics[scale=.3]{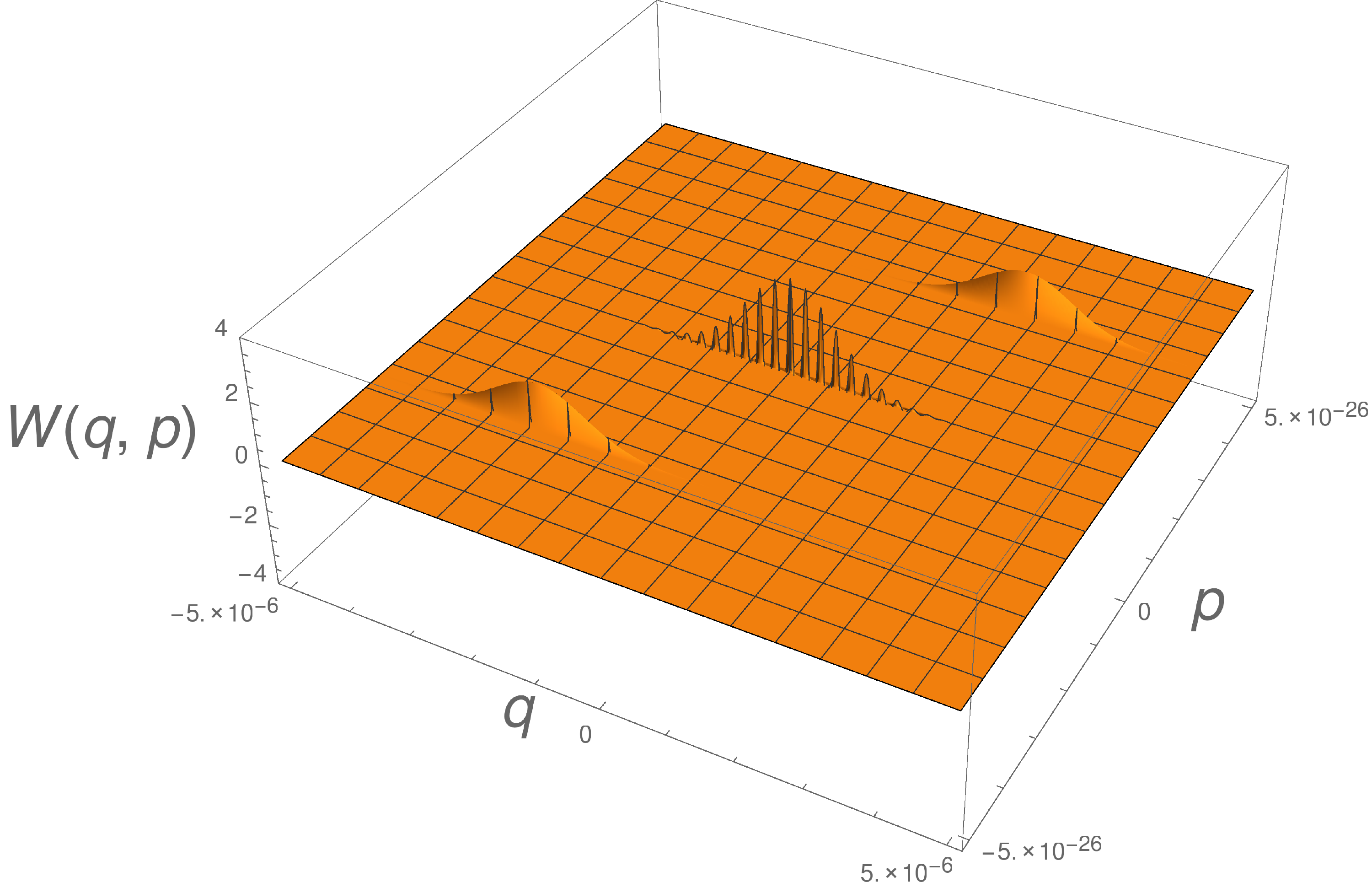}
\caption{\label{fig:epsart} 
Figure shows a plot of the function $W(q,p,t)$ (defined in the text) as a function of $q$ and $p$ 
at t=30$\mu$ sec. The remaining parameters are as mentioned in the caption of Fig. 2. The $W(q,p)$ axis has been rescaled by multiplying by a factor of $10^{-33}$. 
$q$ is displayed in $m$ and $p$ in $kg m/s$. }
\end{figure}

\begin{figure}
\includegraphics[scale=.6]{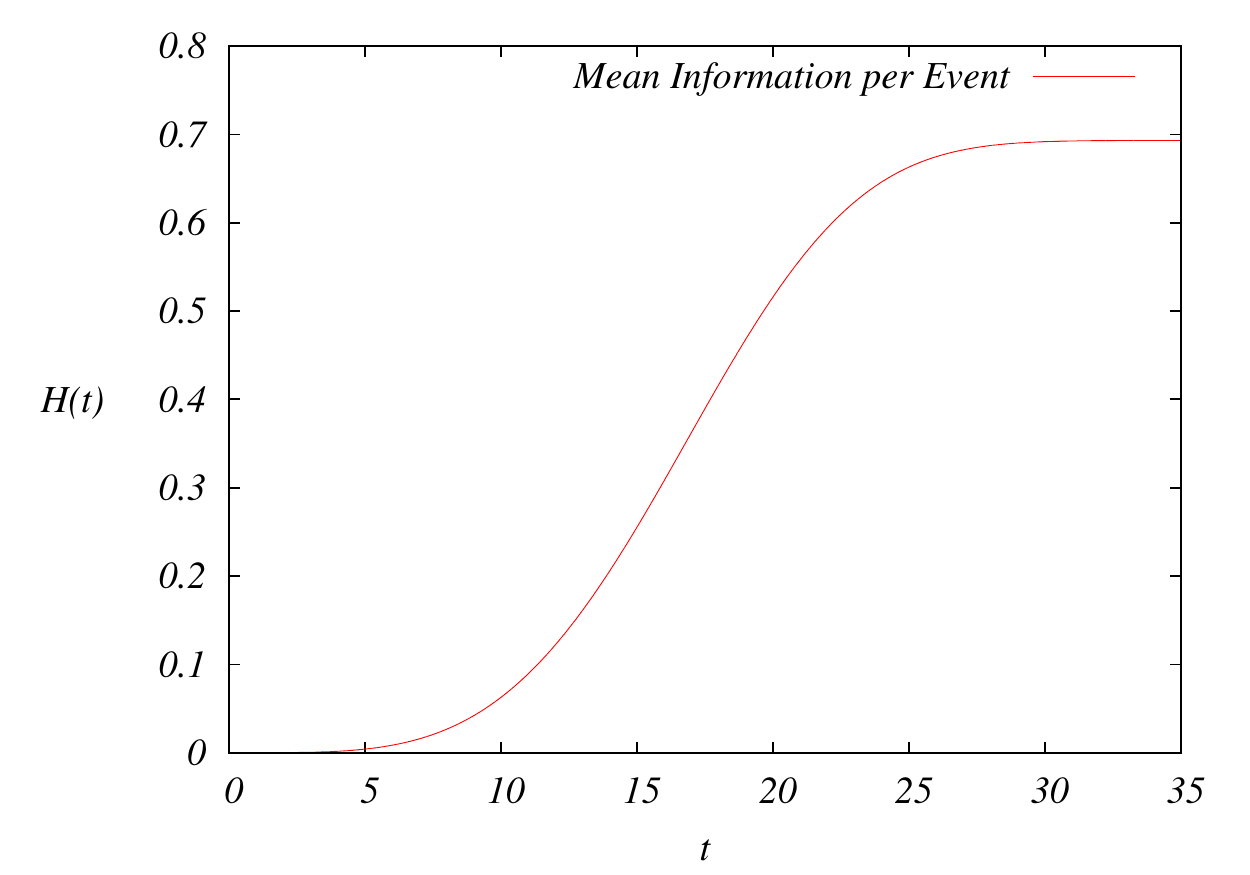}
\caption{\label{fig:epsart} 
Mean information per event $\cal H$ versus time in $\mu$sec. The remaining parameters are as mentioned in the caption of Fig.2. }
\end{figure}

\begin{eqnarray}\nonumber
{\cal H}=\int{P(X) I(X) dX}&=&\label{}\log{2} -\int{P(x) \log{P(x)} dx}\\ \nonumber 
& &+\int{P_+(x) \log{P_+(x)} dx}\\  
& &+\int{P_-(x) \log{P_-(x)} dx}.
\label{mean}
\end{eqnarray}\nonumber
At any time, $\cal H$ cannot of course exceed $S_{ent}$ the entanglement entropy,  
which is the maximum possible information that can be gained about the spin, 
by interrogating the position variable. It follows from Eq.(13) that
${\cal H}$ starts out from zero at $t=0$ and approaches $\log 2$ over a timescale of $\tau_1=10^{-5}
{\rm s}$ (See Fig. 5). The fact that ${\cal H}$ is positive means that we 
do gain some information  about the spin of the atoms by detecting their positions, even before the two wave packets have separated cleanly.
After the wave packets have separated, the coherence between the two wavepackets is still manifest in the Wigner function. 
As is well known, the Wigner function is only a quasiprobability distribution. On coarse graining it becomes positive\cite{kb}
and can be viewed 
as a probability distribution in the phase space of the atomic position.  
This constitutes {\it coarse measurement}.

\section{Conclusion}
\label{}

In this letter we have addressed the Quantum Measurement Process
in the context of the Stern-Gerlach experiment. 
An exact solution of the Schr\"odinger equation
permits us to analyse unitary evolution in an idealised mathematical model 
of the experiment. Coarse Quantum Measurement (CQM) is based on the idea 
that every measurement is done with limited resources of resolution. 
The key conclusion of our analysis is that, the apparent loss of 
unitarity in a quantum measurement 
is a consequence of the coarseness of the experimental probes. Previous literature on coarse measurements 
\cite{kb,jlk,rss} has not applied the idea to understand the classic Stern-Gerlach experiment which
is of great interest as a paradigm for quantum measurement.

In the context of statistical mechanics, the authors of Ref. \cite{jaynes,grandy} note that 
entropy is a subjective notion depending upon the resources available to the experimenter, to distinguish
between statistical states.  
This follows the Bayesian approach to probability theory. 
The view we advocate is very similar in the context of quantum mechanics.
The idea of ``coarse measurement'' is clearly a subjective one. 
Depending on the resources available to the experimenter, the evolution may appear unitary or otherwise. 
Thus, with a high enough resolution one can always detect interference effects. When the interference 
between the two wave packets is detectable, we must conclude 
that the spin is {\it both} up and down simultaneously. 
This does not constitute a measurement of the spin component $\sigma_y$.
In a low resolution  experiment, the interference {\it apparently} gets washed out and we can 
obtain information about the spin. This is the regime of interest in this letter.

Some of the quantum measurement literature concerns itself with von Neuman measurements, which can be regarded
as instantaneous. One talks about ``before'' and ``after'' the measurement, but not during. Exceptions
are weak\cite{weak} and nonideal measurements\cite{home}. 
In weak measurements one tries to continuously extract information from a quantum system causing minimal disturbance using a weak probe,
that does not destroy the interference pattern. In coarse measurements,
one explicitly loses the interference pattern. Regarding non-ideal measurements Ref. \cite{home}
discusses the subtleties in the notion of distinguishability of apparatus
states: even states which are orthogonal in the Hilbert space sense can have considerable spatial overlap.
In contrast, our focus is on how a coarse measurement results in the apparent loss of coherence of the final wavepacket 
in a Stern-Gerlach setup.
As has been emphasized by Ref. \cite{kb}, the coarse measurement approach is conceptually different from the
decoherence paradigm. Decoherence involves interaction with environmental degrees of freedom. Information is lost from the 
system by tracing over the environment. The coarse measurement approach does not invoke new degrees of freedom or new dynamics. 
It is essentially kinematical, dealing with the experimenter's inability to measure or control fine details.

There have been other parallel developments \cite{Namiki,Busch} which address the issue of imperfect measurements.
In \cite{Namiki} the authors 
model the detector as a phase randomizer or dephaser, 
which leads to a mixed state density matrix
starting from a pure state density matrix. 
In \cite{Busch}, the formalism of coarse-graining has been framed   
in a formal mathematical language. 
We go beyond this discussion by providing a physical basis in 
terms of resource limitation. Ref. \cite{Busch} also touches upon the 
issue of non-idealness as in Ref. \cite{home}. 
There has even been a suggestion \cite{Devereux} that the ``reduction of the wavepacket'' happens just when the atom
enters the magnetic field.

In the actual experimental setup for the Stern-Gerlach experiment the atoms are heated in an oven to about $450 K$. At this temperature,
the two spin states of the silver atom are in an incoherent or classical superposition of the two spin states. As a result, the 
interference effects dealt with here will not be visible. 
To see the quantum interference effects discussed here, the internal state of the atom must be in a {\it coherent} superposition of 
spin states. An optical analog of the Stern-Gerlach experiment\cite{ritchie}  
may be a more practical candidate for realizing this experiment.

%*\input \acknowledgement.tex   % input acknowledgement
\section{Acknowledgements}
It is a pleasure to thank Patrick Dasgupta, N. D. Hari Dass, 
Anupam Garg, Chitrabhanu P., Karthik H. S., Gordon Love and Rafael Sorkin  
for stimulating discussions. We thank U. Sinha for drawing our attention to Ref.\cite{home}
 
\appendix
\section{Propagator For The Stern-Gerlach Setup}

We consider a Stern-Gerlach setup with a magnetic field $\boldsymbol{B}=(B_0y,B_0x,0)$. The Hamiltonian for the problem is:
\begin{equation}\tag{A-1}
H=\frac{{p}^2}{2m}-\boldsymbol{{\mu}.{B}}
\end{equation}
with $\boldsymbol{\mu}=g\mu_B\frac{\hbar}{2}{\boldsymbol{\sigma}}$
and the stationary solution satisfies:
\begin{equation}\tag{A-2}
H\psi=E\psi\,\left(\psi=\psi(x,z),\,\psi(x,0)=e^{-\frac{x^2}{2\sigma^2}}\right)
\end{equation}
Restricting to the $x$-$z$ plane by setting $y=0$, the Hamiltonian can be written more explicitly as 
\begin{equation}\tag{A-3}
H=\frac{p_x^2+p_z^2}{2m}+xF\sigma_y
\end{equation}
where, $ F=-g\mu_B\frac{\hbar}{2}B_0$ and if we assume a solution of the form $ \psi(x,z)=\phi(x,z)\,e^{ikz} $ (since this is a propagating wave along z-direction),
Eq.(A-2) in the paraxial approximation reduces to the following:
\begin{equation}\tag{A-4}
i\hbar\left(\frac{k\hbar}{m}\right)\frac{\partial \phi}{\partial z}=-\frac{\hbar^2}{2m}\frac{\partial^2\phi}{\partial x^2}+xF\sigma_y\phi
\end{equation}
if we set $ E=\frac{k^2\hbar^2}{2m}$.

This equation can be identified with the time dependent Schr\"odinger equation, by setting $ t=\frac{zm}{k\hbar} $:
\begin{equation}\tag{A-5}
i\hbar\frac{\partial \phi}{\partial t}=-\frac{\hbar^2}{2m}\frac{\partial^2\phi}{\partial x^2}+xF\sigma_y\phi
\end{equation}
Since the eigenvalues of $ \sigma_y $ are +1 and -1 we get the corresponding components of the spinor $ \phi $ as $ \phi_+ $ and $ \phi_- $, respectively. Thus Eq.(5) reduces to:

\begin{equation}\tag{A-6}
i\hbar\frac{\partial \phi_+}{\partial t}=-\frac{\hbar^2}{2m}\frac{\partial^2\phi_+}{\partial x^2}+xF\phi_+
\end{equation}
\begin{equation}\tag{A-7}
i\hbar\frac{\partial \phi_-}{\partial t}=-\frac{\hbar^2}{2m}\frac{\partial^2\phi_-}{\partial x^2}-xF\phi_-
\end{equation}

To solve these equations we move to an accelerated frame along the x-axis and employ the following transformations in $x$, $t$ and $\phi$, 
which reduce the above equations to a free particle equation for $\tilde{\phi}_\pm(\xi,T)$ where $\xi$ is related to $x$ as follows:
\begin{equation}\tag{A-8}
 x=\xi\pm\frac{1}{2}aT^2\\
\end{equation}
and
\begin{equation}\tag{A-9}
 t=T\\
\end{equation}
Thus we have:
\begin{equation}\tag{A-10}
 \phi_\pm(x,t)=\tilde{\phi}_\pm(\xi,T)\,e^{if(\xi,T)}
\end{equation}
We outline the solution to Eq.(A-6).Substituting Eqs. (A-8), (A-9) and (A-10) in Eq.(A-6) we reduce Eq.(A-6) to a free particle equation and find $f$ and $a$ :
\begin{equation}\tag{A-11}
f=\frac{maT}{\hbar}(\xi+\frac{1}{3}aT^2),\,a=-\frac{F}{m}
\end{equation}
The kernel (propagator) for the free particle problem corresponding to Eq.(A-6) is: 
\begin{equation}\tag{A-12}
\tilde{K}(x,x_i;t)=\sqrt{\frac{m}{2\pi i\hbar t}}\ e^{\,\displaystyle i\frac{m(x-x_i)^2}{2\hbar t}}
\end{equation}
where $x$ is the position at time $t$ and $x_i$ is position at $t_i=0$. 
We can find the propagator for the Hamiltonian under consideration by applying the following transformation.
\begin{equation}\tag{A-13}
K(x,x_i;t)=e^{if(x,t)}\tilde{K}(x,x_i;t)
\end{equation}
which gives Eq.(5) of the main text:
\begin{equation}\tag{A-14}
K^{^{++}}(x,x_i;t) = \sqrt{\frac{m}{2\pi i\hbar t}}\text{exp}\left\{i\left(\frac{m}{2\hbar t}(x-x_i)^2-\frac{Ft}{2\hbar}(x+x_i)-\frac{F^2t^3}{24m\hbar}\right)\right\}
\end{equation}
The solution to Eq.(A-6) can be cast as follows:
\begin{equation}\tag{A-15}
\phi_+(x,t)=\int_{-\infty}^{\infty}K^{++}(x,x_i,t)\,\phi_+(x_i,0)\,dx_i
\end{equation}
After solving Eq.(A-13) we get the final solution for $\phi_+$. We employ the same procedure to find $\phi_-$. The solutions are:
%\begin{subequations}
%\begin{empheq}{align}
%\phi_+(x,t)=\sqrt{\frac{m\sigma}{(m\sigma^2+i\hbar t)\sqrt{\pi}}}\ \text{exp}\left\{\,\displaystyle-\frac{m(12x^2+\frac{a^2}{\hbar}(4im\sigma^2-\hbar t)t^3+\frac{12axt}{\hbar}(-2im\sigma^2+\hbar t))}{24(m\sigma^2+i\hbar t)}\right\}\\
%\phi_-(x,t)=\sqrt{\frac{m\sigma}{(m\sigma^2+i\hbar t)\sqrt{\pi}}}\ \text{exp}\left\{\,\displaystyle-\frac{m(12x^2+\frac{a^2}{\hbar}(4im\sigma^2-\hbar t)t^3-\frac{12axt}{\hbar}(-2im\sigma^2+\hbar t))}{24(m\sigma^2+i\hbar t)}\right\}
%\end{empheq}
%\end{subequations}
\begin{equation}\tag{A-16}
 \resizebox{1\hsize}{!}{$\phi_+(x,t)=\sqrt{\frac{m\sigma}{(m\sigma^2+i\hbar t)\sqrt{\pi}}}\ \text{exp}\left\{\,\displaystyle-\frac{m(12x^2+\frac{a^2}{\hbar}(4im\sigma^2-\hbar t)t^3+\frac{12axt}{\hbar}(-2im\sigma^2+\hbar t))}{24(m\sigma^2+i\hbar t)}\right\}$}
\end{equation}
\begin{equation}\tag{A-17}
 \resizebox{1\hsize}{!}{$\phi_-(x,t)=\sqrt{\frac{m\sigma}{(m\sigma^2+i\hbar t)\sqrt{\pi}}}\ \text{exp}\left\{\,\displaystyle-\frac{m(12x^2+\frac{a^2}{\hbar}(4im\sigma^2-\hbar t)t^3-\frac{12axt}{\hbar}(-2im\sigma^2+\hbar t))}{24(m\sigma^2+i\hbar t)}\right\}$}
\end{equation}

In general, one can consider a further evolution beyond the region where the magnetic field is present and consider free evolution which 
leads to solutions of the form given below: 

\begin{equation}\tag{A-18}
 \resizebox{1\hsize}{!}{$|\phi_+(x,t)|^2=\sqrt{\frac{m^2\sigma^2}{(m^2\sigma^4+\hbar^2 t^2){\pi}}}\ \text{exp}\left\{\,\displaystyle
-\frac{m^2[x-\frac{1}{2}a{t_1}^2-at_1(t-t_1)]^2}
{(m^2\sigma^4+\hbar^2 t^2)}\right\}$}
\end{equation}
\begin{equation}\tag{A-19}
\resizebox{1\hsize}{!}{$|\phi_-(x,t)|^2=\sqrt{\frac{m^2\sigma^2}{(m^2\sigma^4+\hbar^2 t^2){\pi}}}\ \text{exp}\left\{\,\displaystyle
-\frac{m^2[x+\frac{1}{2}a{t_1}^2+at_1(t-t_1)]^2}
{(m^2\sigma^4+\hbar^2 t^2)}\right\}$}
\end{equation}

where $t_1$ is the amount of time spent by the atom in the magnetic field and $t$ is the total time of evolution.
\section{Suppression of off diagonal elements of the Wigner matrix due to coarse graining}

The Wigner matrix is of the form:
\begin{equation}\tag{B-1}
\mathbf{W}(q,p)=\left( \begin{array}{cc}  W_{++}(q,p) & W_{+-}(q,p) \\
 W_{-+}(q,p)& W_{--}(q,p) \end{array} \right)
\end{equation}

Explicitly, for instance, we have:
\begin{equation}\tag{B-2}
 \resizebox{1.0\hsize}{!}{$W_{++}(q,p)= \frac{1}{2\pi \hbar}\text{exp}\left\{ {-\frac{\frac{4p^2t^2}{m^2}-\frac{4pt(at^2+2q)}{m}+(at^2+2q)^2+\frac{4p^2\sigma^4}{\hbar^2}-\frac{8ampt\sigma^4}{\hbar^2}+\frac{4a^2m^2t^2\sigma^4}{\hbar^2}}{4\sigma^2}} \right \}$} 
\label{pp}
\end{equation}
\begin{equation}\tag{B-3}
W_{+-}(q,p)= {\frac{1}{2\pi\hbar}}\text{exp}\left[-\left\{\frac{(pt-mq)^2}{m^2\sigma^2}+\frac{p^2\sigma^2}{\hbar^2}+\frac{iat(pt-2mq)}{\hbar}\right\}\right]
\label{pm}
\end{equation}

Notice that $W_{+-}(q,p)$ oscillates on a spatial scale $d=\frac{\hbar}{2mat}=\frac{\hbar}{2Ft}$. 
The coarse grained Wigner matrix $\overline{\mathbf{W}} (q,p)$ has elements
\begin{equation}\tag{B-4}
\overline{W}_{\alpha\beta} (q,p)=\frac{1}{\Delta \delta}\int_{-\Delta/2}^{\Delta/2} du\int_{-\delta/2}^{\delta/2} dv 
{W}_{\alpha\beta} (q+u,p+v),
\label{wigneraverage}
\end{equation}
with $\Delta$ and $\delta$, the pixel size in position and momentum respectively.
The numerically generated plots show how the offdiagonal terms
${W}_{+ -}(q,p)$ and ${W}_{-+}(q,p)$ on coarse graining average to zero due to the presence of the oscillatory term $e^{iq\frac{2\pi}{d}}$,
where $d$ is the spatial scale of oscillation. We finally get the following diagonal form for the coarse grained Wigner matrix:

\begin{center}
$\left(\begin{array}{ccc}
\overline{W}_{++}(q,p) & 0\\
& & \\
0 & \overline{W}_{- -}(q,p)
\end{array}\right)$
\end{center}
%%\includegraphics[scale=0.3]{W0.pdf}
%%\caption{\label{fig:epstart}
%%Figure shows a plot of the Wigner function $W(x,p,t)$ as a function of $x$ and $p$ at $t=0$. 
%%We have chosen $\hbar =1$, $m=3$, $a=3$ and $\sigma=1$.} 
%%\end{figure}
For instance, for $q=10^{-6}m$ and $p= 0 kgm/s$ and $t=3\times10^{-5}s$ we get the following form for the coarse 
grained Wigner matrix, for the realistic experimental parameter values for a typical Stern Gerlach setup. 
\begin{equation}\tag{B-5}
\overline{\mathbf{W}}(q,p)=\left( \begin{array}{cc}  5.7\times10^{-2} & 0 \\
 0 & 1.6\times10^{-5} \end{array} \right)
\end{equation}

\biboptions{sort&compress}

\bibliographystyle{elsarticle-num} 
\bibliography{supurnareferences}





\end{document}